\documentclass[superscriptaddress,showpacs,twocolumn,amssymb,aps,prl]{revtex4}
\usepackage{graphicx,dcolumn,bm,amsmath}

\newcommand{\fig}[1]{Fig.~\ref{#1}}

\newcommand{\eeq}{ \end{equation} }
\newcommand{\beq}{ \begin{equation} }

\newcommand{\bhu}{ \hat{\bf u} }

\newcommand{\br}{ {\bf r} }

\begin{document}

\title{How to capture active particles}

\author{A. Kaiser}
\affiliation{Institut f\"ur Theoretische Physik II: Weiche Materie,
Heinrich-Heine-Universit\"at D\"{u}sseldorf, 
Universit{\"a}tsstra{\ss}e 1, D-40225 D\"{u}sseldorf,
Germany}
\author{H. H. Wensink}
\email{wensink@lps.u-psud.fr}
\affiliation{Institut f\"ur Theoretische Physik II: Weiche Materie,
Heinrich-Heine-Universit\"at D\"{u}sseldorf, 
Universit{\"a}tsstra{\ss}e 1, D-40225 D\"{u}sseldorf,
Germany}
\affiliation{Laboratoire de Physique des Solides, Universit\'{e} Paris-Sud 11, B\^{a}timent 510, 91405 Orsay Cedex, France}
\author{H. L\"{o}wen}
\affiliation{Institut f\"ur Theoretische Physik II: Weiche Materie,
Heinrich-Heine-Universit\"at D\"{u}sseldorf, 
Universit{\"a}tsstra{\ss}e 1, D-40225 D\"{u}sseldorf,
Germany}

\date{\today}

\begin{abstract}
For many applications, it is important to catch collections of 
autonomously navigating microbes and man-made microswimmers
in a controlled way.
Here we propose an efficient trap to collectively capture
 self-propelled colloidal rods. By means of computer simulation
in two dimensions, we show that a static chevron-shaped wall  represents an optimal boundary for a trapping device. Its catching efficiency can be tuned by varying the
opening angle $\alpha$ of the trap. For increasing $\alpha$,
there is a sequence of three emergent states corresponding to 
partial, complete, and no trapping.  A trapping `phase diagram' maps out the trap conditions
under which the capture of self-propelled particles at a given  density is rendered optimal.
\end{abstract}

\maketitle

One of the key survival strategies of human beings over the ages is their ability to catch
animals. An efficient way to capture mammals, fish and birds is ``trapping'', i.e.\ to release a 
device in a populated zone  which then irreversibly
attracts and stores the prey. While the methods for capturing  (macroscopic) animals have been well-optimized by now, 
the corresponding problem in the micro-world, namely catching microbes, is much more challenging due to the
strongly reduced  nanometric size of the trap. The possibility to trap autonomously navigating
microorganisms in a controlled way provides fascinating options to prevent or cure microbial 
contamination \cite{contamination1,contamination2} and to concentrate microbes near externally imposed patterned surfaces \cite{chaikin}.  Similar applications can be envisaged for 
 man-made microswimmers, i.e. 
artificial particles which are actively propagating due to an internal ``motor".
Examples  include catalytically driven Janus particles \cite{Golestanian,Baraban,Bocquet}, colloids 
with artificial flagella \cite{Bibette,Ignacio} and vibrated granulates \cite{AransonRMP,Ramaswamy}.  
Lithographic techniques have been employed to confine, control and steer the motion of microbes and artificial microswimmers \cite{Kaehr,Bechinger}.  The use of lithographic nanopatterns   has advanced significantly in recent years \cite{microfluidic,Clement}  and has opened up numerous possibilities to sort particles \cite{hulme,Baskaran}, to rectify their motion \cite{chaikin,ReichhardtPRL},
and to design building blocks of micro-machines 
\cite{DiLeonardoPNAS,SokolovPNAS,Angelani}. Despite  the experimental evidence there is little fundamental understanding of trapping phenomena in systems of active particles. 
This provides impetus for investigating the minimum requirements for designing  efficient microbial traps and for unravelling the physical mechanisms responsible for collective trapping active particles.

Here, we propose an efficient scenario for capturing self-propelled rods by subjecting them to a static confining boundary of variable shape. 
The  collective dynamics of the rods in confined geometry is explored by computer simulation using a two-dimensional,  particle-resolved model for self-propelled colloidal rods.
The presence of a boundary dramatically changes the collective dynamics of the rods, which in bulk show a strong propensity to form swarms, and induces collective self-trapping near the boundary. 
We show that a chevron-type (V-shaped) boundary represents the ideal geometry for an efficient microbial trap.  The opening angle $\alpha$ of the chevron plays a key role in determing the self-trapping efficiency of the set-up. For increasing $\alpha$, the following sequence of non-equilibrium stationary states emerges:
partial trapping, complete trapping, and no trapping. The transition from partial to complete trapping occurs 
smoothly at a lower critical opening angle while the complete trapping state abruptly terminates 
at an upper critical opening angle. The trapping phenomenon highlighted in this study is a collective 
effect which is driven by the many-body dynamics of self-motile rods in confinement. We show that the sharp
transition from complete to incomplete trapping at higher critical opening angles has an appropriate 
system-size scaling which allows a classification similar to a true thermodynamic phase transition.
The results of our simulation study are relevant for both two and three-dimensional systems of self-propelled particles. 
In three spatial dimensions the optimal shape of the trap would be a circular cone. 

The model consists of a collection of 
$N$ active rods with length $\ell$ which experience a constant
 self-motile force $F_{a}$ directed along the main rod axis of each
 rod. Due to solvent friction the particles move in the overdamped
low-Reynolds number regime, while interacting with the other particles and the
boundary by steric forces only \cite{Wensink}. The latter are implemented by
discretizing the rod length into $n$ spherical segments and imposing a repulsive
Yukawa potential  between the
segments of any two rods. The total potential
between a rod pair $\{\alpha, \beta \}$ with orientational unit
vectors $\{ \bhu _{\alpha}, \bhu _{\beta} \}$ and centre-of-mass
distance ${\bf \Delta r}_{\alpha \beta}$ is then given by 
$U_{\alpha \beta} = (U_{0}/n^{2}) \sum_{i=1}^{n}\sum_{j=1}^{n}\exp
[-r_{ij}^{\alpha \beta} / \lambda]/r_{ij}^{\alpha \beta} $ where
$U_{0}>0$ defines the amplitude, $\lambda$ the screening length and
$r_{ij}^{\alpha \beta} = |{\bf \Delta r}_{\alpha \beta} + (l_{i}
\bhu_{\alpha} - l_{j} \bhu_{\beta})|$ the distance
between segement $i$ of rod $\alpha$ and $j$ of rod $\beta$ ($\alpha
\neq \beta $) with $l_{i} = d(i-1)$, $i \in [1,n]$
denoting the segment position along the symmetry axis of the
rod. The number of segments $n$ per rod is chosen such as to guarantee an intrarod
segment-segment distance $d=\ell/(n-1) \leq \lambda$ to prevent rods from overlapping. A trap is introduced as a static boundary with a
prescribed shape and contour length $\ell_{T}$. Particle-trap interactions are
implemented by discretizing the trap boundary into $n_{T} = \lfloor \ell_{T}/d \rceil $ equidistant
segments each interacting with the rod segments via the same Yukawa potential. Mutual rod collisions generate apolar nematic alignment which favor active liquid crystalline order, e.g. swarms, if the rod concentration exceeds a certain critical value. The boundary potential mimics a hard wall and imparts 2D planar order with rods pointing favorably perpendicular to the local wall normal.

 In the overdamped regime,
the equations of motion for the positions
and orientations of the
$N$-particle system in the absence of thermal
noise emerge from a balance of the friction, interaction and
self-propulsion forces and torques acting on each rod $\alpha$:
\begin{eqnarray}
{\bf f }_{\cal T} \cdot \partial_{t} \br_{\alpha}(t) &=&  -\nabla_{\br_{\alpha}}
 U(t) +  F_{a} \bhu_{\alpha}(t), \nonumber \\
{\bf f}_{\cal{R}} \cdot \partial_{t} \bhu_{\alpha}(t) &=&
-\nabla_{\bhu_{\alpha}} U(t),
\label{eom}
\end{eqnarray}
in terms of the total potential energy  $U=(1/2)\sum_{\alpha, \beta (\alpha \neq
  \beta)} U_{\alpha \beta} + \sum_{\alpha ,T} U_{\alpha T}$
with $U_{\alpha T}$ the potential energy of rod $\alpha$ with the
trap, and friction tensors ${\bf f}_{\cal T} =
f_{\parallel}\bhu\bhu + f_{\perp}({\bf I} - \bhu\bhu)$ and ${\bf
  f}_{\cal R} = f_{\cal R}{\bf I}$ (with ${\bf I}$ the 2D unit tensor). The Stokesian friction factors
$f_{\parallel},f_{\perp},f_{\cal R}$ correspond to the  parallel, perpendicular
and rotational degrees of motion of a single rod and depend solely on the rod
aspect ratio, defined as $a = \ell/\lambda $
\cite{tirado}. The self-propulsion speed of a single  rod is 
simply $v_{0}=F_{a}/f_{\parallel}$ independent of  time $t$ which is conveniently expressed in units $\tau = f_{\parallel}\lambda/F_{a}$, i.e. the time a free rod requires to traverse a distance equalling its length $\ell$. 
\begin{figure}
\begin{center}
\includegraphics[clip=,width= 0.6\columnwidth ]{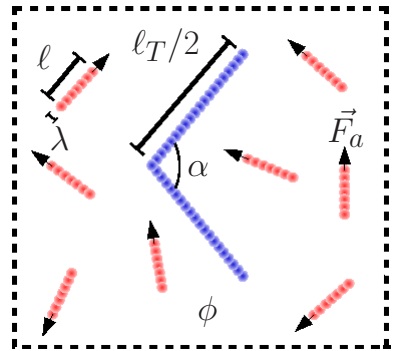}
\caption{ \label{f1} Sketch of a
  system of self-motile rods 
with aspect ratio $a= \ell/\lambda$ and axial active force  $F_{a}$ at bulk density $\phi$ subjected to a static chevron-type
  trap with contour length $\ell_{T}$ and variable opening angle $\alpha$.  The macroscopic system consists of periodic replicas with boundaries indicated by the dotted lines. } 
\end{center}
\end{figure}
\begin{figure*}
\begin{center}
\includegraphics[clip=,width= 1.9\columnwidth ]{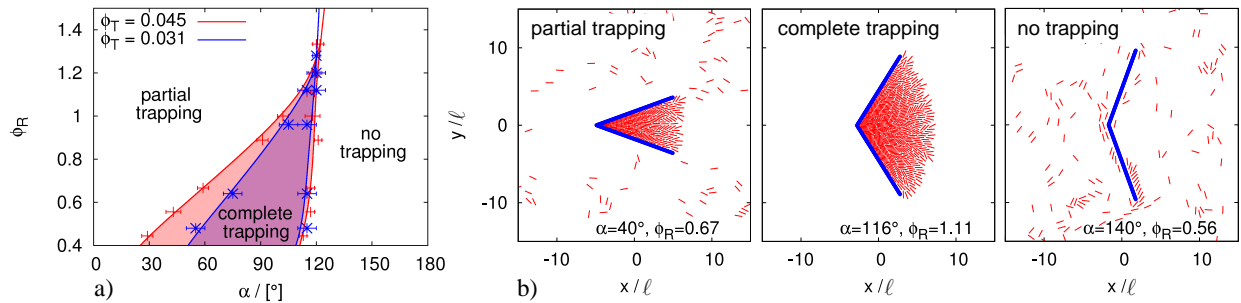}
\caption{ \label{f2} (a) `Phase diagram' marking the three different trapping states for a chevron-type boundary with length $\ell_{T}=20\ell$; no trapping at large trap angle $\alpha$, complete trapping at medium angle and partial trapping at small angle upon variation of the reduced rod packing fraction $\phi_{R}=\phi/\phi_{T}$.  Phase boundaries are shown for two different values of $\phi_{T}$; the area fraction occupied by the trap. The region of complete trapping is bounded by a triple point at larger rod concentration beyond which a smooth transition from no trapping to partial trapping occurs. (b) Snapshots depicting the three stationary states for the case $\phi_{T}=0.045$.} 
\end{center}
\end{figure*}
We simulate rods with aspect ratio $a=10$ in a rectangular simulation box with area
$A$ and periodic boundary conditions in both Cartesian
directions.  A particle packing fraction is defined as $\phi = N\sigma/A$ with
$\sigma= \ell\lambda + \lambda^{2}\pi/4$ the effective area of
a single rod.  In bulk these systems will spontaneously form polar nematic swarms
with number density fluctuations being much stronger than for the
(passive) equilibrium system at a  comparable bulk density \cite{Narayan}.
  
Let us now subject the swimmers  to a V-shaped trap  with
contour length $\ell_{T} = 20 \ell$ and variable opening angle $0^{\circ} <
\alpha < 180^{\circ}$ (see \fig{f1}). In the macroscopic limit,  the system can be interpreted 
as a reservoir of microswimmers exposed to an equidistant array of
mutually independent static traps. The area fraction occupied by the traps is angle-dependent with a maximum value given by $\phi_{T} =
(\ell_{T}^{2}/8A)$, which fixes  the number of rods via $N=(\ell_{T}^{2}/8 \sigma)(\phi/\phi_{T})$. The trap area fraction shall be constrained to values below 0.1
in order to guarantee the traps to be completely independent of each other within
the typical range of bulk rod packing fractions  $0 < \phi < 0.1$ consider here. 
Trapped particles are identified by imposing a simple immobility criterion such that a particle is considered trapped if its velocity $v$ is below some threshold velocity $v^{\ast}$ ($v=|{\bf v}|<v^{\ast}$) during a long time interval of at least $t^{\ast} = 25\tau$. 
This criterion suffices to disregard any particles that may be transiently anchored to the trap boundary. The number fraction of trapped rods defines the trapping efficiency $x_{T}(t)=N_{T}(t)/N$ and its long-time limit $x_{T}^{(0)}=\lim_{t\to \infty } x_{T}(t)$  ($0<x_{T}^{(0)}<1$) can be used to discern various stationary states.

\begin{figure}
\begin{center}
\includegraphics[clip=,width= 1\columnwidth ]{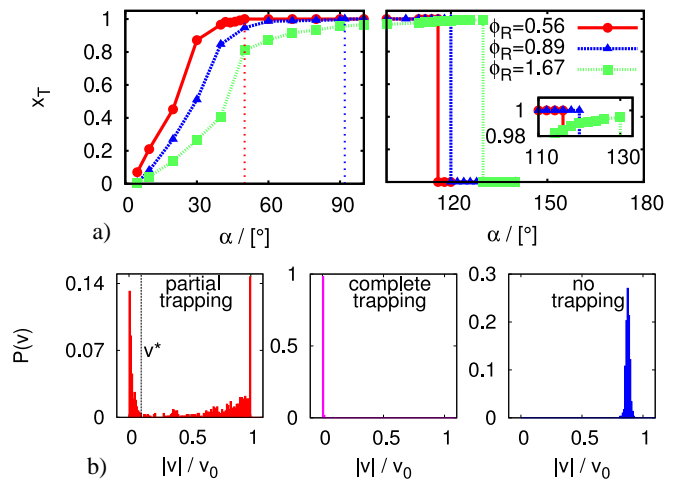}
\caption{ \label{f3}  (a) Number fraction of trapped self-propelled rods $ x_{T}^{(0)} = N_{T}(\infty)/N$ as a function of the opening angle $\alpha$ for three different reduced rod packing fractions $\phi_{R}$.  The jumps around $\alpha=120^{\circ}$ signal  `first-order' phase transitions from no trapping to complete trapping.  The dotted vertical lines in the left panel locate the transitions from complete to partial trapping upon decreasing $\alpha$. Inset: behaviour of $x_T$ near the critical trapping angle $\alpha \approx 120^{\circ}$.   (b) Histograms of the normalized distribution $P(v)$ of particle velocities $v=|{\bf v}|$ corresponding to the three different states.  } 
\end{center}
\end{figure}

\fig{f2}a represents an overview of the collective trapping states that arise upon varying the two main system variables, the rod packing fraction  $\phi$, and opening angle
$\alpha$.  The trapping `phase diagram' exhibits three stationary regimes. First, for large $\alpha$ no trapping occurs as the cusp is too wide to efficiently capture a significant fraction of particles in the system. Second, below a certain critical angle a sharp transition towards  {\em complete
trapping} occurs. This state is characterized  by the formation of a large monocluster in the cusp of the trap which contains {\em all particles} in the system.  Upon further decreasing $\alpha$ a third region is entered  corresponding to {\em partial trapping}. In this regime, the effective rod-trap collision cross-section is insufficient to trap all the particles present in the system,  and a substantial portion of the rods remains mobile even at large $t$.   
The transition lines demarcating the various states in the diagram can be inferred from the evolution of the `order parameter' $x_{T}^{(0)}$, i.e., the number fraction of trapped rods, as a function of the trap angle $\alpha$, as shown in \fig{f3}a.   Upon departing from the flat-wall limit ($\alpha \sim 180^{\circ}$) a sharp discontinuity occurs where $x_{T}^{(0)}$ jumps from zero to unity which signals a transition from no trapping to complete trapping.  For sharper cusps (smaller $\alpha$) a second, continuous transition from complete to partial trapping can be located at the point where $x_{T}^{(0)}$  starts dropping smoothly below unity.  The choice of the velocity cut-off $v^{\ast}=0.1v_{0}$, used in determining the trapping order parameter $x_{T}^{(0)}$,  is robust as can be   justified  from the shape of the velocity histograms $P(v)$ in \fig{f3}b where the markedly separated peaks  provide an unambiguous distinction between immobile (trapped) particles and freely mov
 ing
 ones. From the histograms one can infer that the untrapped particles in the partial trapping state predominantly move freely with a maximum velocity $v_{0}$ whereas those in the no-trapping state are slowed down significantly due to collisions with the trap boundary.

 It is worth noting from \fig{f2}a that the transition from no trapping to complete trapping appears rather insensitive to the rod concentration as well the area fraction occupied by trap. Generically,  complete trapping is possible only if the trap angle does not exceed a typical threshold value $\alpha \approx 120^{\circ}$.  Moreover, by defining a reduced rod density $\phi_{R}=\phi/\phi_{T}$ the triple point is rendered virtually independent of the trap area fraction and attains a universal value $\phi_{R}^{\ast} \approx 1.3$. This suggests that the window of stability for complete trapping, as demarcated by  the rod density $\phi^{\ast}$ at the triple point, can be systematically tuned by changing the number of traps per area and/or the contour length $\ell_{T}$ of the boundary. Although the state of complete-trapping is strictly suppressed at rod densities $\phi > \phi^{\ast}$, the trapping efficiency $x_{T}$ still shows a marked jump from zero to nearly 100\% if the angle
 drops below about 120$^{\circ}$. This indicates that the V-shaped boundary continues to be a powerful trapping device at larger rod densities.

\begin{figure}
\begin{center}
\includegraphics[clip=,width= 0.9\columnwidth ]{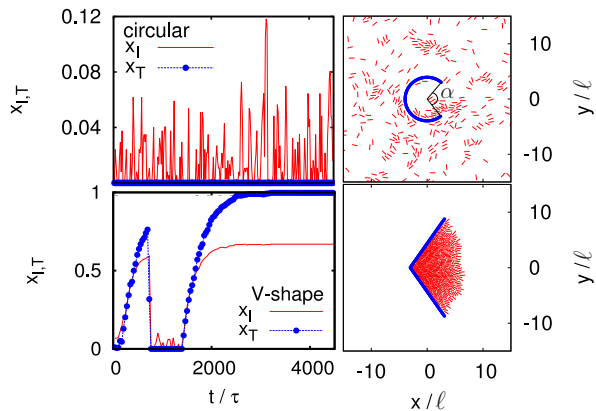}
\caption{ \label{f4}   Comparison of the number fraction of particles $x_{I}$ located within the trap area and the fraction of immobile, trapped particles $x_{T}$ as a function of time $t$  for both  V-shaped and circular traps with equal trap area $A_{T} = 52 \ell^{2}$,  aperture angle $\alpha =110^{\circ} $ and reduced rod packing fraction $\phi_{R} = 0.89$. } 
\end{center}
\end{figure}

To ascertain  whether the  chevron  is indeed the optimal
shape we compare its trapping efficiency with that of a circular trap  with identical trap area $A_{T} = 52 \ell^{2}$ and opening angle $\alpha = 110^{\circ}$. A time series of the fraction of trapped particles reveals a distinct difference between the two trap shapes (\fig{f4}).  While the V-shape induces a fast intake of particles into the trap surface leading up to a trapping efficiency of almost 100\% (complete trapping) at large times, the circular one fails to capture a significant fraction of particles over time. The rounded shape of the circular trap  does not facilitate particle-wall anchoring but instead forces clusters of particles to slide collectively along the inner contour of the circular trap. This leads to a process whereby the rods collectively enter and leave the interior of the trap, as indicated by the `bursts' in the number fraction of particles $x_{I}$ inside the
trap in \fig{f4}a. The rhythmic nature of the process is reflected by a peak in the
power spectrum at a characteristic frequency which translates into a typical life time $212 \tau$ of a rod cluster inside the circle trap.  The number fraction of trapped (i.e. immobile) rods $x_{T}(t)$, however, remains practically zero throughout the sampled time interval in stark contrast to what is observed for  the chevron trap.

In conclusion, while there is a wealth of knowledge about trapping passive particles,  e.g. colloids in optical tweezers or atoms in a
Paul trap, there is considerable less understanding
of capturing self-propelled particles. Here
 we have shown that active rods are
efficiently self-trapped by a static chevron-type boundary. At the kink of the chevron,
 active particles are forced to oppose each other and form a jammed
cluster which acts as a nucleus for capturing more particles  until the whole
trap area is filled. The opening angle of the trap plays a crucial role and its variation leads to three different emergent states corresponding
to no trapping, complete trapping and partial trapping.
A trap boundary which is rounded on the length scale of the particle extension 
is incapable of capturing particles over time.
It is therefore essential that the trap boundary possesses at least one sharp
cusp (whose size should be comparable to the rod length) which facilitates a fast build-up of clustered rods.
The self-trapping mechanism is present in three spatial dimensions where the optimal
trapping boundary is represented by a circular cone.
We emphasize that the trapping mechanism proposed here is a collective, non-equilibrium effect which is conceptually different from trapping due to an equilibrium external
force as represented by e.g. a deep potential well. In the equilibrium situation,
collective trapping can be understood on the single-particle level. Moreover, 
geometric details of the trap boundary are usually irrelevant for understanding the global behaviour in contrast to the boundary-induced self-trapping mechanism advanced in this study. 

The trapping phenomena presented here should be verifiable
in experiments on rod-shaped bacteria \cite{cisnerosgoldstein} or driven polar granular
rods \cite{kudrolli} exposed to  geometrically structured boundaries \cite{Bechinger, microfluidic,Clement}. This set-up could, for instance, be exploited as an efficient purification device to manipulate and remove contaminating microbes.
Future investigation will be aimed at comparing  different types of propulsion mechanisms
\cite{stark,gompper} and the effect of the associated hydrodynamic flow fields generated
by the collective particle motion in the presence of a static trap boundary \cite{Dhont}.
We believe, however, that the hydrodynamic interactions will neither change the
microscopic mechanism underpinning collective trapping nor the topology of the phase diagram presented here.
\acknowledgments We thank A. Menzel for helpful discussions. This work is supported by the DFG within SFB TR6 (project section D3).

\bibliographystyle{apsrev}

\bibliography{refs}

\end{document}